\documentclass[%
 reprint,
 amsmath,amssymb,
 aps,
]{revtex4-2}

\usepackage{graphicx}
\usepackage{dcolumn}
\usepackage{bm}

\begin{document}

\preprint{APS/123-QED}

\title{Effective Magnetic Switching of THz Signal in Planar Structured Spintronic Emitters}

\author{Evgeny A. Karashtin$^{1,2,}$}
 \email{eugenk@ipmras.ru}
\author{Nikita S. Gusev$^{1}$}%
\author{Maksim V. Sapozhnikov$^{1,2}$}
\author{Pavel Yu. Avdeev$^{3}$}
\author{Ekaterina D. Lebedeva$^{3}$}
\author{Anastasiya V. Gorbatova$^{3}$}
\author{Arseniy M. Buryakov$^{3}$}
\author{Elena D. Mishina$^{3}$}

\affiliation{$^{1}$Institute for Physics of Microstructures RAS, Nizhny Novgorod,  Russia}
\affiliation{$^{2}$Lobachevsky State University of Nizhny Novgorod, Nizhny Novgorod, Russia}
\affiliation{$^{3}$ MIREA – Russian Technological University, Moscow, Russia}%

\date{\today}

\begin{abstract}
We demonstrate an efficient swithcable spintronic THz emission from a Co(2nm)/Pt(2nm) bilayer split into a grating of micro- or nanostripes in applied external magnetic field of the order of 10mT.
The THz signal is effectively emitted if the samples are magnetized perpendicular to the stripes direction and decreases upon a 90-degree rotation of a sample. The maximal 27-fold ratio of amplitude change is achieved for the sample with a $2 \mu m$ width of stripes. For the samples with easy magnetization axis directed perpendicular to stripes the THz signal is linearly polarized in the direction of the stripes and abruptly switches during the sample re-magnetization process. If the easy axis is along the stripes the THz source is almost switched off in a zero magnetic field.
\end{abstract}

\maketitle

\graphicspath{{Figure/}}

\section{\label{Intro} Introduction}
In the last decade, promising sources of THz radiation based on nanostructures consisting of nano-thick ferromagnetic layers (FM) having boundaries with nano-thick layers of a non-magnetic (usually heavy) metal (NM) have been actively studied and promoted \cite{CBul2021spintronic,10.1063/5.0051217,10.1063/5.0080357,Leitenstorfer2023, THz_sources,yurasov2024magnetorefractive}. The minimal effective structure consists of two layers: FM/NM. Numerous studies have shown that such structures can act as broadband THz emitters, comparable in efficiency to semiconductor analogues, ZnTe, semiconductor-based antennas (see for review \cite{Kampfrath2013,CBul2021spintronic,Leitenstorfer2023,vedmedenko20202020,10.1063/5.0080357, Wu2021}). Besides they have the sufficient advantage because THz emission can be manipulated by magnetic field in this case \cite{Kong2019}.

When such a system is irradiated with a femtosecond high-intensity optical pulse, a bunch of hot electrons is generated. In a magnetic material, these hot electrons are spin-polarized, which results in a short pulse of “pure” spin current flowing from the FM to the NM. The spin current induces an electric (charge) current in the NM via the inverse spin Hall effect. Such a non-stationary electric current emits a short and broadband electromagnetic wave pulse corresponding to the terahertz frequency range \cite{CBul2021spintronic}. Using this technology, THz fields of 1.5 MV/cm have already been achieved \cite{Kampfrath2023}.The highest efficiency of optical-to-THz conversion was obtained for double \cite{10.1063/5.0167151,Buryakov2023} and triple \cite{10.1063/5.0167151,Kampfrath2023} layer structures. 

Structures in which in-plane anisotropy is introduced make it possible to completely control the polarization of the emitted THz wave by changing magnitude of the magnetic field \cite{Khusyainov2021,10.1063/5.0185251}. Controlling amplitude by changing magnitude of the magnetic field while polarization remains constant has also been demonstrated in a spin valve type structure consisting of two layers of iron separated by a 4 nm thick layer of platinum with one of the iron layers being attached to antiferromagnetic IrMn layer \cite{Fix2020}. In those structures, both Fe layers inject spin current into Pt. Changing magnetization from parallel to antiparallel results in antiparallel or parallel charge current from opposite sides of the structure, thus providing minimal (zero) or maximal THz amplitude. The IrMn layer also emits THz waves when a spin current is injected into it from the neighboring iron layer, but an order of magnitude less efficiently than Pt \cite{Chen2018, Mikhaylovskiy2023}. As a result, in \cite{Fix2020} the ratio of THz signal intensities in the “on” and “off” modes of the device is approximately 15. In \cite{Shahriar2023}, modulation of THz intensity by up to 87\% is achieved in On-Chip Wave-guided SiNx/Fe/Pt/SiNx structure. THz signal modulation is a primary area of research in THz communication, focusing on key parameters such as materials and structures, modulation depth, modulation speed, and others \cite{ma2019modulators, liu2021terahertz}. Electrical modulation, while having shown good results, requires further improvement to achieve deeper and faster modulation. Optical modulation, on the other hand, already demonstrates excellent characteristics in these areas, but there is still room for improvement, making it a promising direction for further scientific research and development \cite{ma2019modulators}.

In this study, we propose a structure that provides a novel mechanism for controlling the amplitude of a THz signal from a spintronics emitter using an external magnetic field. We used a standard cobalt-platinum bilayer and etched a grating in it with a sub-wavelength period throughout the entire depth of the bilayer. This array acts as a THz antenna, where a magnetic field applied perpendicularly or parallel to the stripes allows the amplitude of the THz wave to be modulated with a high modulation depth. The efficiency of amplitude modulation has been studied depending on the width and period of the stripes, and a theoretical justification for the observed dependencies has been given.

The paper is organized as follows. In Section~\ref{Setup} we describe preparation of the samples and the experimental measurement technique. Section~\ref{Res} compiles the experimental data for THz signal from all samples. In Section~\ref{Disc}, the results of measurements are discussed and appropriate theory is verified showing reasonable fitting parameters.

\section{\label{Setup} Methods and Samples}

\begin{figure}[tb]
\centering{\includegraphics[width=\hsize]{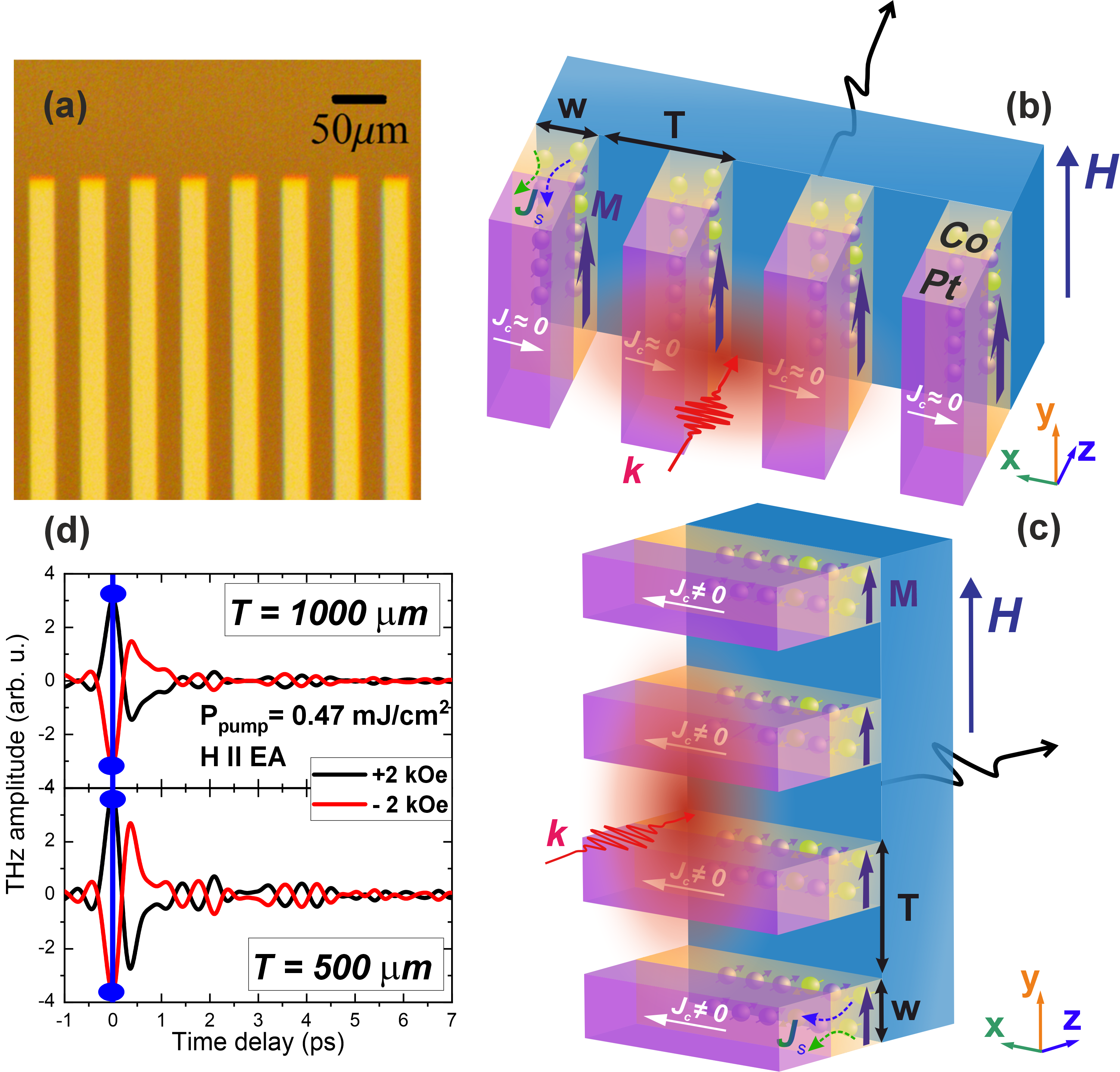}}
\caption{\label{fig1} (a) Optical photograph of a $50 \mu m$ period grating. (b,c) Schematic representation of current (Jc) and amplitude of generated THz signal for the magnetic field applied along the stripes or perpendicular to them, respectively. (d) Typical time dependence of terahertz field emitted from the Co(2)/Pt(2) for the grating with $1000 \mu m$ and $500 \mu m$ period ($500 \mu m$ and $250 \mu m$ stripe width, respectively).
}
\end{figure}

The spintronic terahertz emitters are Co(2)/Pt(2) bilayer films (hereafter, thickness is in nm, order of layers is from substrate) grown on quartz substrates by magnetron sputtering. An external magnetic field of approximately $1 kOe$ was applied during growth in order to induce a uniaxial anisotropy in the sample plane. Details of growth technology may be found elsewhere \cite{Fraerman2002}. The $1cm \times 1cm$-size gratings with microns to hundreds of microns period ($4, 50, 100, 300, 500, 1000 \mu m$) and $1/2$ filling factor are then made from these films by optical lithography (Figure~\ref{fig1}(a)). Two series of samples were made. In the first series, the stripes were lithographed along the magnetic anisotropy axis of the original film. In the second series, the stripes were lithographed across the anisotropy axis of the original film. We also make a sample with $8 \mu m$ period and $3/4$ filling factor ($6 \mu m$ stripe width). 


Sample planar structure characterization was performed by optical microscopy. Magnetic properties were checked by longitudinal magnetooptical Kerr rotation (MOKE) investigation with the use of He-Ne laser with the wavelength of $630$ $nm$.
THz emission measurements were conducted by using a standard THz time-domain spectroscopy method. A Ti:Sapphire laser with an amplifier system generated pulses with a central wavelength of $800$ $nm$, repetition rate of $3 kHz$, and pulse duration of $35 fs$. Samples of spintronic THz emitters were placed in a constant external magnetic field of up to $2 kOe$, created by an electromagnet.
A magnetic field was applied along the Y-axis of the laboratory coordinate system (Figure~\ref{fig1}(c)). This study considered two geometries for magnetic field application: magnetizing samples along an easy axis $H \parallel \text{E.A.}$ and perpendicular to it. These geometries are achieved by a 90-degree rotation of a sample in the laboratory coordinate system. This is shown schematically in Figure~\ref{fig1}(b,c) for structures in which the easy axis is aligned along the direction of the stripes. These pictures also illustrate the mechanism of THz generation due to the inverse spin-Hall effect (ISHE) in both configurations (arrows indicate the directions of the spin current $Js$ injected from the ferromagnetic layer into the layer of non-magnetic Pt and the charge current $Jc$ generated as a result of ISHE).

The generated THz signal was focused onto a nonlinear optical crystal, ZnTe, using a system of parabolic mirrors. To detect the THz signal, a standard electro-optical sampling technique was employed with a setup that included the ZnTe detector, a quarter-wave plate, a Wollaston prism, and a balanced photodetector. The polarization of the pump and probe beams was aligned parallel to the X-axis of the laboratory coordinate system (Figure~\ref{fig1}(c)) and the ZnTe axis (-110). The laser spot size on the samples was $3 mm$ at the FWHM level, corresponding to a pump energy density of $0.47 mJ/cm^{2}$. A wire-grid polarizer installed in the THz radiation path, along with the THz polarization-sensitive ZnTe, enabled the analysis of the THz radiation polarization. All measurements presented in this work were performed for the Ex component of the THz field. A more detailed description of the THz detection setup can be found in Ref. \cite{Buryakov2022}.

\section{\label{Res} Experimental results}
\begin{figure}[tb]
\centering{\includegraphics[width=\hsize]{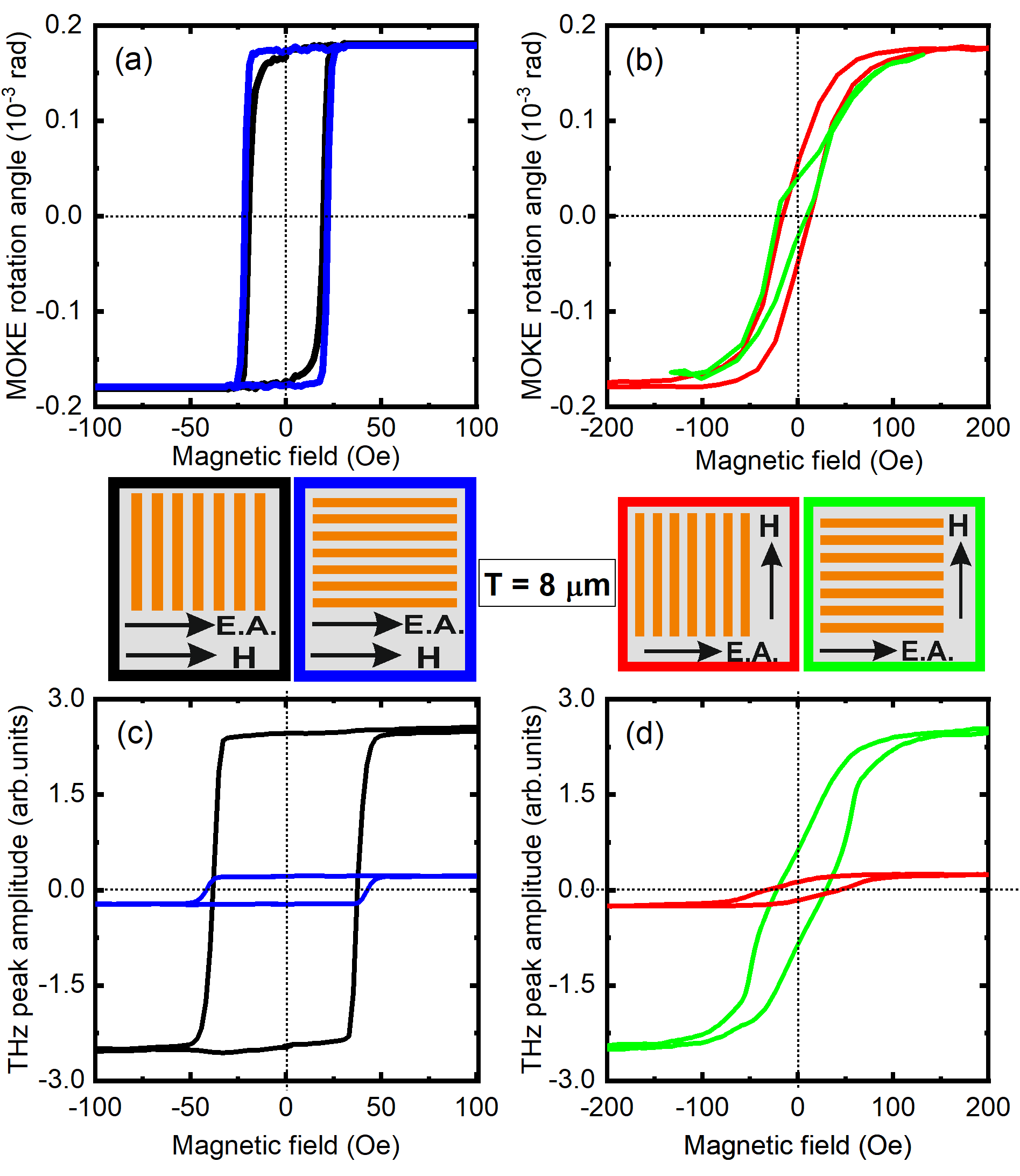}}
\caption{\label{fig2} MOKE hysteresis loop (a) along the easy axis (E.A.) and (b) perpendicular to the easy axis (E.A.) for the sample with a $8 \mu m$ period grating and a $6 \mu m$ stripe width and different orientation of easy axis with respect to the stripe direction, as shown in the inset. (c,~d) THz loops for the corresponding geometry of samples and measurement.}
\end{figure}

All samples are verified via the MOKE measurements along and perpendicular to the easy axis. Typical MOKE curves are shown in Figure~\ref{fig2}(a,b) for the grating with $8 \mu m$ period and $6 \mu m$ wire width.
Measurements of magnetization loops show that lithography of the lattice does not change the direction of magnetic anisotropy in the sample; it retains the same direction that it had in the original film. Therefore, in the first series of samples (as introduced in Section~\ref{Setup}), the easy axis of magnetic anisotropy is directed along the lattice stripes, and in the second series of samples, across the lattice. When magnetized along the anisotropy axis, the samples have a rectangular hysteresis loop, the coercivity field is in the range of 20-100~Oe. When magnetized across the easy magnetization axis, the samples demonstrate an inclined hysteresis loop with a residual magnetization in the range of 0.2-0.3 of saturation magnetization and a saturation field in the range of 100-300~Oe. Non-zero remanent magnetization in this case is due to a polycrystalline structure of the sample.

We perform time dependence of THz field amplitude for the setup when the grating is irradiated by an optical pump from front (sample) and back (substrate) side. The main quantitative difference is the magnitude of signal due to different absorption coefficients for optical and THz signal in a quartz substrate. 
Note that the results are qualitatively the same both for front and back side. 
Typical result for irradiation from front side is shown in Figure~\ref{fig1}(d) for $1000 \mu m$ and $500 \mu m$ grating period with $1/2$ filling factor and close geometry. The THz hysteresis loops are constructed based on these measurements as follows. The THz signal is measured at a delay position of 0 ps, which corresponds to the peak maximum (as shown in Figure~\ref{fig1}(d)). Then, at this position, a magnetic field is applied, and the corresponding THz amplitude is recorded for each value of the magnetic field. Typically, the THz signal polarized in the direction perpendicular to the magnetic field applied to the sample (with the magnetic field along the y-direction) is recorded. We designate this usual signal polarization as X-polarization.


The main experimental results of the article are presented in Figure~\ref{fig2}(c,d) and can be briefly formulated as follows. The shape of the dependence of the THz generation amplitude on the applied magnetic field repeats the shape of the magnetization loop of the samples and is determined by the orientation of the field relative to the magnetic axis of anisotropy. Conversely, the maximum generation amplitude observed in saturation depends on the direction of the external magnetic field relative to the orientation of the lattice stripes. 
An observed discrepancy in hysteresis loop width for MOKE and THz measurements may be attributed to the processes such as sample heating by laser beam \cite{Buryakov24_APLM}.

\begin{figure}[b]
\centering{\includegraphics[width=\hsize]{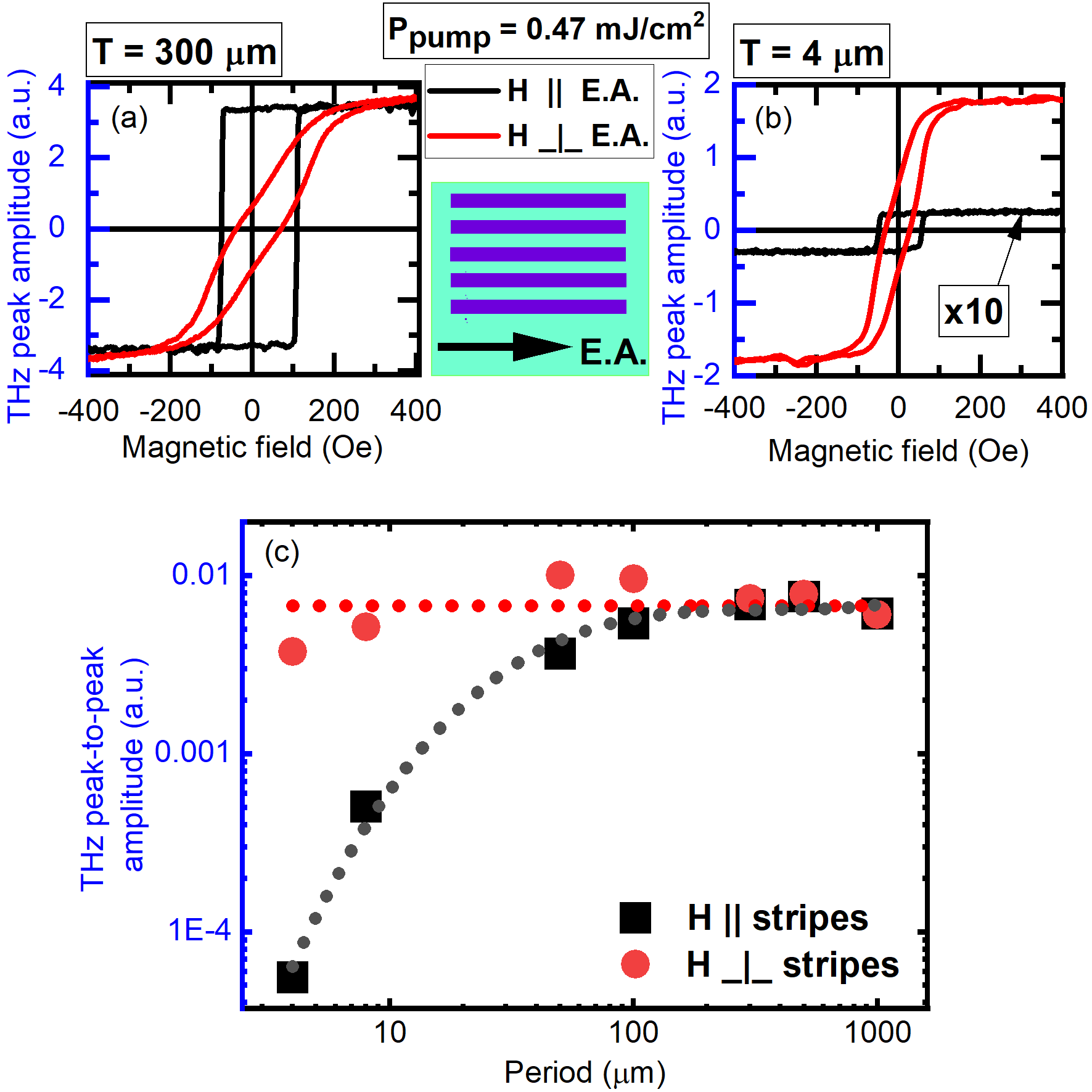}}
\caption{\label{fig4} (a,b) THz hysteresis loop for the lattices with 1/2 filling factor and grating period of $300 \mu m$ and $4 \mu m$, correspondingly. The stripes are along E.A.; the magnetic field is applied along (black line) or perpendicular (red line) to the easy axis of magnetic anisotropy. (c) Dependence of THz amplitude in a saturating magnetic field applied along (black squares) or perpendicular (red circles) to the stripes. The lines are guide for eyes.}
\end{figure}

We showed above that the shape of THz loop depends only on the E.A. direction with respect to the applied magnetic field and is almost insensitive to the direction of stripes.
Typical THz hysteresis loops for large ($300 \mu m$) and small ($4 \mu m$) period are shown in Figure~\ref{fig4}(a,b) for the case when E.A. is along the stripes. One can note that the THz signal in a saturating magnetic field ($400 Oe$) is almost independent of the field direction for grating with large period, whilst demonstrating a strong (approximately 27-fold) decrease upon a 90-degree rotation from the H.A. (perpendicular to E.A.) direction to the E.A. one.

The THz peak-to-peak amplitude in a saturating magnetic field applied along the stripes or perpendicular to them for different grating period is shown in Figure~\ref{fig4}(c). Note that since the direction of E.A. strongly influences the loop shape leaving the THz signal in a saturating field almost intact Figure~\ref{fig4}(c) collects all the data for samples with different directions of magnetization easy axis. This data is now sorted by the direction of the magnetic field with respect to the stripes (a $2/3$ scaling factor is used for the sample with $6 \mu m$ stripe width and $8 \mu m$ period). It is evident that if the magnetic field is oriented across the gratings, the magnitude of the THz signal in saturation weakly depends on the grating period $T$. A minor decrease for small periods may be attributed to the fact that sufficient power is emitted into side diffraction maxima which are not collected by the parabolic mirror in our setup. Conversely, if the magnetizing field is oriented along the gratings, then a decrease in the grating period leads to a significant decrease in the THz signal. In other words, for gratings with large periods ($1000, 500, 300 \, nm$), turning the saturating field relative to the gratings does not lead to a change in the amplitude of THz generation. Meanwhile, for gratings with a small period ($4 nm$), a 90-degree turn of the saturating field from perpendicular to parallel direction relative to the grating stripes leads to a drastic 27-fold decrease in the generation amplitude.

\begin{figure}[tb]
\centering{\includegraphics[width=0.7\hsize]{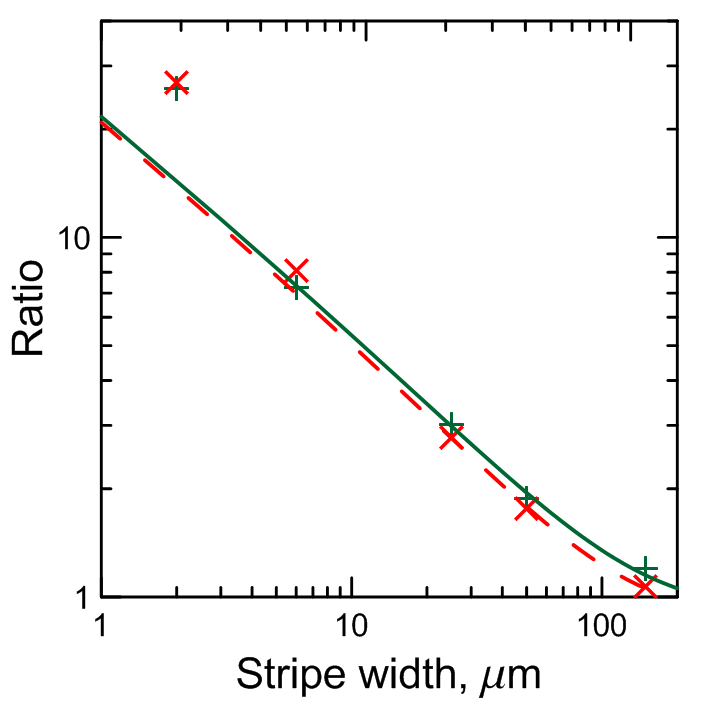}}
\caption{\label{fig7} THz signal ratio versus stripe width for front side (pluses) and back side (crosses) measurment. The lines represent theoretical approximation (solid line for front side, dashed line for back side).}
\end{figure}

Changes in the amplitude of THz generation during rotation of the magnetic field relative to the grating depending on the grating period are summarized in Figure~\ref{fig7} as the ratio of corresponding THz signals (in logarithmic scale).
One can see that the ratio decreases as the stripe width increases, but quite slowly. Similar dependencies are observed for front side and back side measurements. We include the results obtained from the structure with a $8~\mu m$ period (Figure~\ref{fig2}(c,d)) and 3/4 filling factor into Figure~\ref{fig7}. Contrary to our expectations, the absolute THz signal from these structures is not greater than from the structures with 1/2 filling factor (Figure~\ref{fig4}(c)). However the ratio corresponds to theoretical fit (described below) based on their stripe width quite well. 

\begin{figure}[t]
\centering{\includegraphics[width=0.8\hsize]{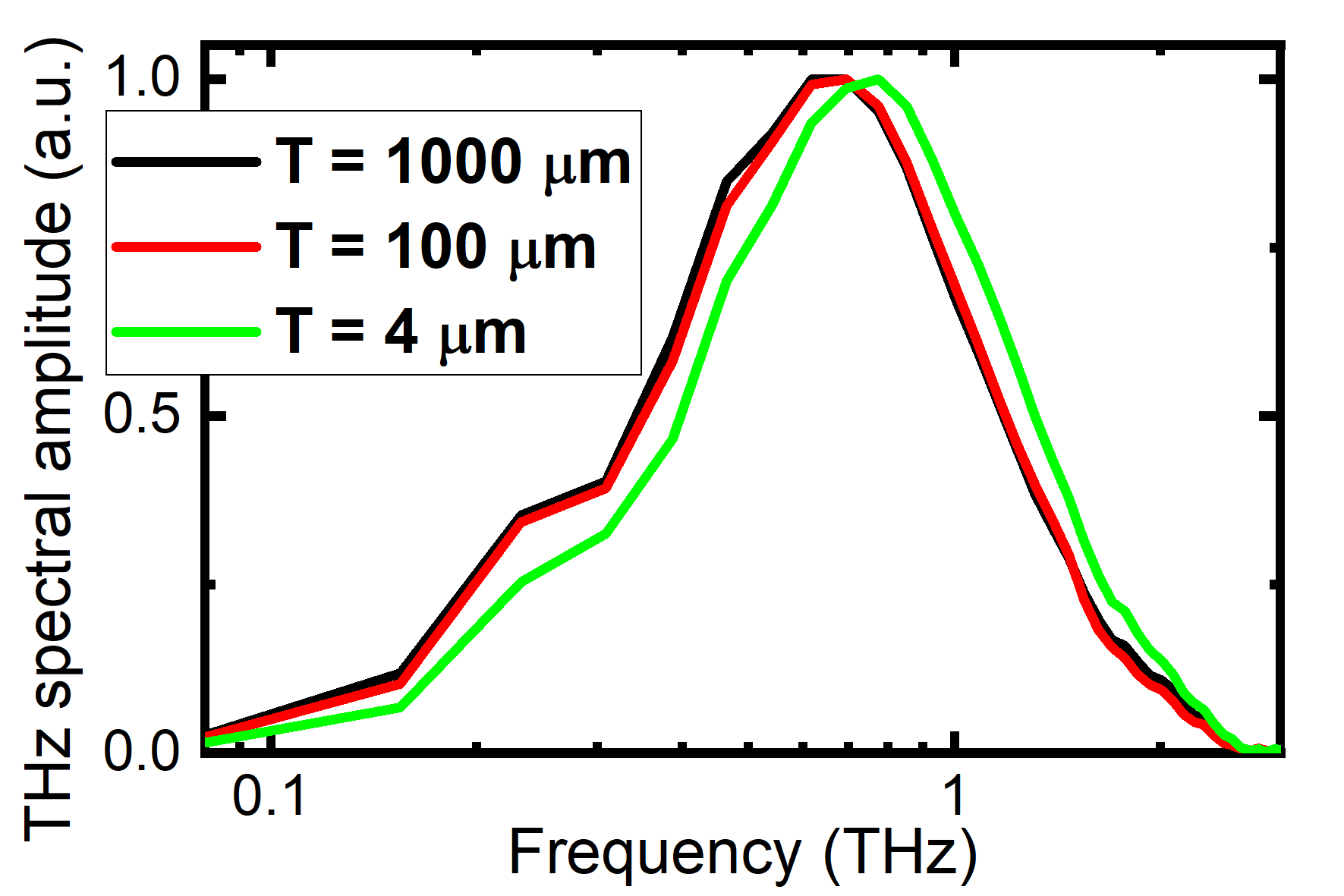}}
\caption{\label{fig3} Normalized spectra of THz signal for different stripe periods. A saturating magnetic field of $400Oe$ is applied perpendicular to the stripes.}
\end{figure}

Spectral properties of the THz signal obtained from the stripes with different representative periods in a saturating magnetic field applied perpendicular to the stripes are shown in Figure~\ref{fig3}. A weak shift of the radiation bandwidth to the high frequency range is observed for small grating period. This shift is absent for $100 \mu m$ period and is roughly $10\%$ (or $0.07 THz$) for $4 \mu m$ period. This may be explained by the fact that very small size of stripes as emitters compared to wavelength becomes important. This slightly suppresses emission for long wavelength while for small wavelength (high frequency) the grating still works as a sum of rather big rectangles.

We aslo measure the other linear polarization (which we denote as Y-polarization) of THz signal (see Supplementary Materials) where the polycrystalline structure of the sample is even more pronounced. The main feature of these measurements is that for narrow stripes the Y-polarized signal is very small when the magnetic field is applied perpendicular to the stripes. This matches the fact that these samples poorly emit THz signal when magnetized along the stripes. Thus almost one linear polarization of the THz signal is emitted well from narrow stripes; the amplitude of this signal can be effectively controlled by the direction of applied magnetic field with respect to the stripes orientation.



\section{\label{Disc} Discussion}


We develop a simple theoretical model in order to explain the experimental data. The model is described in detail in Supplementary Materials; here we briefly point out its main features and pass to the final result for the THz signal ratio which is then used to fit the experimental data.

For the sample magnetized perpendicularly to the stripes, the electric current provided by ISHE flows along the stripes which are supposed to be very long (we neglect charge accumulation at the stripes' ends). We suppose that the power injected into a unit square of the emitter from magnetic system $p_i$ is constant. The electric field emitted by such system is then proportional to the electric current density averaged over the square:
\begin{equation}\label{Eq_jpar}
\left<j_{\parallel}\right> \sim \sqrt{\frac{p_i}{2 \rho h}}.
\end{equation}
where $\rho$ is resistivity, $h$ is the thickness of the stripe. 

\begin{figure}[b]
\centering{\includegraphics[width=\hsize]{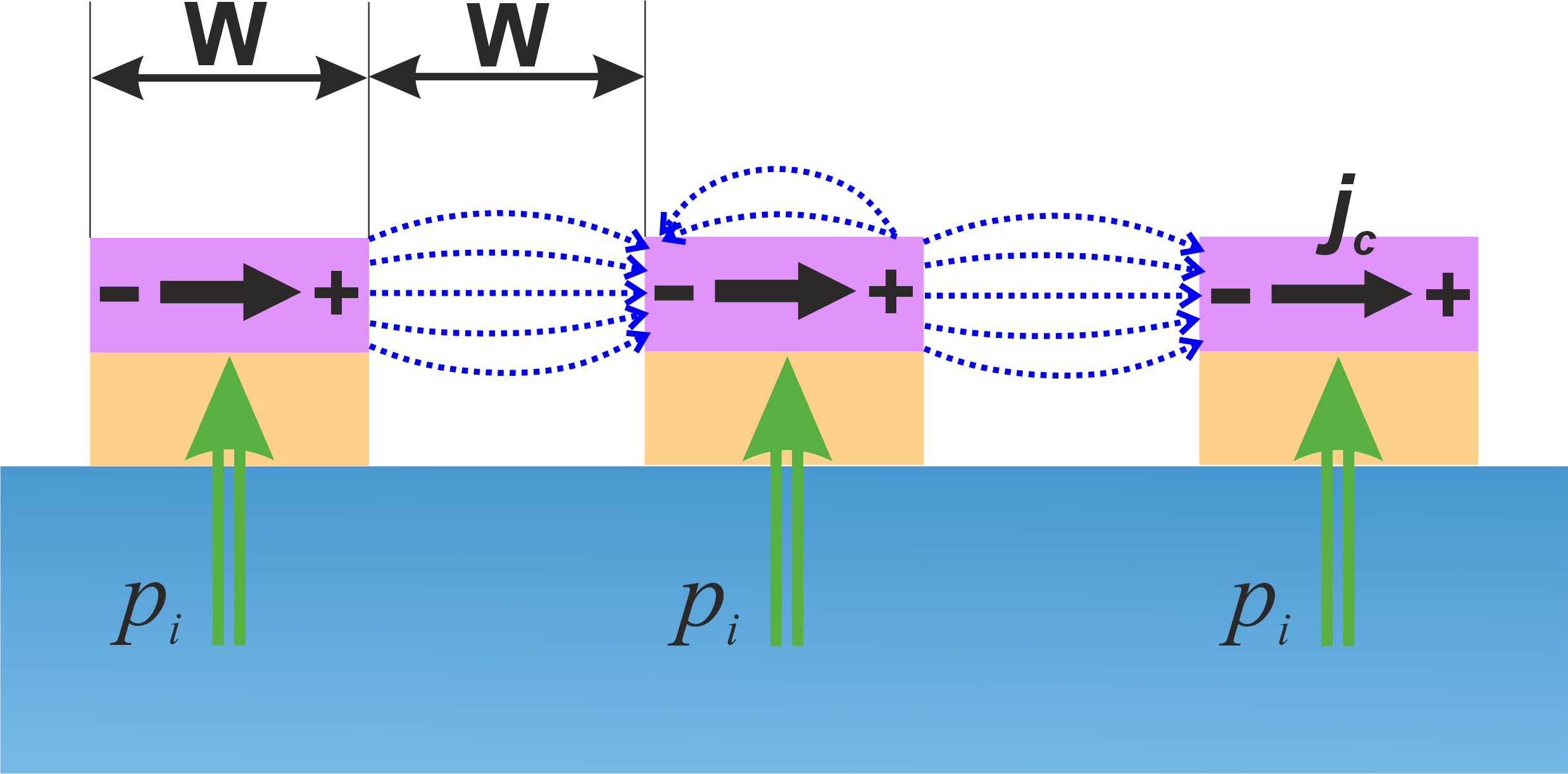}}
\caption{\label{fig8} A schematic view of stripes cross section. The power $p_i$ is injected from a Co ferromagnet (sand color) to a Pt heavy metal (purple color) of a bilayer structure lying on an optically transparent quartz substrate (blue). The electric charges accumulate at the Pt boundaries and create an electric field which leads to energy accumulation.}
\end{figure}

A different case occurs if the sample is magnetized along the stripes: the current caused by ISHE flows perpendicular to the stripes which leads to charge accumulation (Figure~\ref{fig8}).
In order to determine the current in this case we ought to calculate the high-frequency conductance taking into account active resistance determined by film conductivity and reactance mainly determined by electrical capacitance of the system. We may roughly estimate it by supposing that the charge is conducted at very thin wires with the diameter equal to the layer thickness $h$.
We then calculate the energy power accumulated at such a charged edge in the system of stripes and taking into account power absorption due to the wire resistance obtain an estimation of the electric current electric current density averaged over the square in this case as:
\begin{equation} \label{Eq_j}
\left<j_{\perp}\right> \sim \sqrt{\frac{p_i}{2 h \left(\rho + \frac{2}{\omega} \frac{h}{w} Log\left(\frac{L}{w}\right)\right)}},
\end{equation}
where $L$ and $w$ are the characteristic grating width and the stripe width, $\omega$ is the frequency of electric current.
The THz signal ratio follows from (\ref{Eq_jpar}) and (\ref{Eq_j}):
\begin{equation} \label{Eq_ratio}
    Ratio = \frac{\left<j_{\parallel}\right>}{\left<j_{\perp}\right>} = \sqrt{1 + \frac{h}{L}\frac{2}{\rho \omega} \frac{L}{w} Log\left(\frac{L}{w}\right)}.
\end{equation}

The results of experimental data fit with the use of (\ref{Eq_ratio}) are shown in Figure~\ref{fig7} by solid (dashed) line for front (back) side configuration. Note that we use data for relatively big ($> 10 \mu m$) stripe width for the fit. The characteristic grid coherence length $L$ is determined from the fit and is equal to $264 \mu m$ for front side measurements and $184 \mu m$ for back side ones. These values are of the same order; the difference may be attributed to sensitivity of this parameter to small deviations of experimental values caused by noise. One can note that although the real grid size is of the order of $1cm$ we obtain $L$ of the order of THz signal wavelength. This is governed by the fact that equation (\ref{Eq_ratio}) is obtained in the quasistatic approximation which works well for distances less than the wavelength. The other fitting parameter is the stripe effective resistivity $\rho$. We evaluate it from our fit as $\rho = 1.6 \cdot 10^{-7} \Omega \cdot m$ which is the same for measurements from both sides. Considering the structure as a parallel connection of Pt  and Co conductors with the same cross section and resistivity equal to $10.6 \cdot 10^{-8} \Omega \cdot m$ and  $6.2 \cdot 10^{-8} \Omega \cdot m$ respectively \cite{Quantities} we arrive at the structure resistivity equal to $8.4 \cdot 10^{-8} \Omega \cdot m$ which is approximately two times smaller than obtained from our fit. This is of the same order of value; the difference is explained by the fact that thin films usually have greater resistivity than bulk material.

It is worth noting that the model (and the experiment) is almost not sensitive to the distance between stripes unless this distance is of the order of value of the stripe width $w$. This is explained by a logarithmic dependence of the energy stored in the accumulated charge on this distance which leads to a small constant correction under the root in (\ref{Eq_ratio}). Indeed, the point put for $6 \mu m$ stripe width in Figure~\ref{fig7} is measured for the structure with $8 \mu m$ period and 3/4 filling factor. It corresponds to the fit quite well. On the other hand, the ratio value equal to $27$ obtained for the $2 \mu m$ stripe width is slightly greater than that predicted by Equation~\ref{Eq_ratio} (approximately $15$) which may be attributed to the fact that we do not take into account some effects such as the circuit inductance in our rough model. This makes the THz signal ratio even greater.


\section{Conclusion}
In summary, we show that the planar grating that consists of stripes made of typical ferromagnet / heavy metal spintronic terahertz emitter may act as a THz source which intensity can be effectively controlled by application of an external magnetic field: the source emits THz signal well if magnetized perpendicular to the stripes direction and much worse if magnetized along the stripes. The ratio of THz signals is approximately $27$ (almost two times greater than that for a structure with two ferromagnetic layers divided by a relatively thick $4nm$ Pt interlayer \cite{Fix2020}) for the stripe width $2 \mu m$ which is easily achieved by optical lithography. Although the overall THz signal amplitude obtained from our samples is less than from solid films by a filling factor our method may give a greater signal than that suggested in literature \cite{Fix2020} because of a much smaller total thickness of layers that we use.
Our samples emit the THz signal linearly polarized along the stripes; thus we straightforwardly intergate a linear polarization THz antenna into the spintronic emitters. 

The THz signal from the samples with crystalline anisotropy perpendicular to the stripes direction almost abruptly switches from one THz field polarization direction to an opposite one upon a remagnetization process which gives two controllable THz signal states. On the other hand, for the samples with crystalline anisotropy parallel to the stripe direction the THz field goes from a maximum in a saturating field to a very small value in zero field thus allowing to controllably achieve three states of THz signal. This opens up a possibility to create either two- or three-state controllable sources for possible use in THz logic devices \cite{Ortiz-Martinez2020, Wang2022}.

\begin{acknowledgments}
This work was supported by Center of Excellence ``Center of Photonics'' funded by The Ministry of Science and Higher Education of the Russian Federation, contract 075-15-2022-316. THz measurement experiments were supported by the RSF (Grant No. 23-19-00849).
\end{acknowledgments}


\bibliography{apssamp}

\end{document}